\documentclass[12pt]{article}
\usepackage{amssymb,amsmath}
\usepackage{epsfig}

\def\appendix#1{
  \addtocounter{section}{1}
 \setcounter{equation}{0}
  \renewcommand{\thesection}{\Alph{section}}
 \section*{Appendix \thesection\protect\indent \parbox[t]{11.715cm} {#1}}
  \addcontentsline{toc}{section}{Appendix \thesection\ \ \ #1}
  }

\newcommand{\newsection}{
\setcounter{equation}{0}
\section}

\newcommand{\eq}[1]{\begin{equation} #1 \end{equation}}
\newcommand{\ar}[1]{\begin{eqnarray} #1 \end{eqnarray}}
\newcommand{\tr}{\mathop{\mathrm{tr}}\nolimits}

\def\const{{\rm const}}
\def\e{{\,\rm e}\,}
\def\d{\partial}
\def\D{\delta}
\def\dd{^{\dagger}}
\newcommand{\br}[1]{\left( #1 \right)}
\newcommand{\vev}[1]{\left\langle #1 \right\rangle}
\newcommand{\rf}[1]{(\ref{#1})}
\newcommand{\non}{\nonumber \\*}
\def\N{${\cal N}=4$ }
\def\es{\varepsilon}
\def\ep{\epsilon}
\def\th{\theta}
\def\dx{\dot{x}}
\def\st{\sqrt{g^2N}}
\newcommand{\bc}[1]{\left. #1 \right|_{\d D}}
\def\cl{X_{\rm cl}}
\def\ph{\varphi}
\def\cir{{\rm circle}}
\def\bs{\bar{\sigma}}
\def\ds{d\mathfrak{s}}

\title{
Supersymmetric Wilson loops}

\author{K. Zarembo\thanks{{\tt Konstantin.Zarembo@teorfys.uu.se}. Also at
ITEP, B.~Cheremushkinskaya 25, 117259 Moscow, Russia}
\\~~\\
Department of Theoretical Physics\\ 
Uppsala University \\
Box 803, SE-751 08 Uppsala, Sweden 
}

\begin{document}           

\maketitle 
               
\abstract{I construct 1/16, 1/8 and 1/4 BPS
Wilson loops in \N supersymmetric Yang-Mills theory and argue that
expectation values of 1/4 BPS loops do not receive quantum corrections.
At strong coupling, non-renormalization of supersymmetric Wilson loops
implies subtle cancellations in the partition function
of the AdS string with special boundary conditions. The cancellations
are shown to occur in the semiclassical approximation.
}

\newsection{Introduction}

The holographic duality of \N supersymmetric Yang-Mills theory
and type IIB string theory on $AdS_5\times S^5$ 
\cite{Maldacena:1998re,Gubser:1998bc,Witten:1998qj,Aharony:1999ti}
is one of the few examples in which the long suspected
 equivalence of strings and planar diagrams
of the large-$N$ limit \cite{'tHooft:1974jz}
can be established  on the quantitative level.
However, this equivalence is not very explicit in the AdS/CFT correspondence. 
The string picture is useful
in the strong coupling regime of SYM, when all planar diagrams should be
taken into account. Resummation of all planar graphs is equivalent
to solving the large-$N$ limit, which is impossible 
 in an interacting field theory, such as
\N SYM. 

Though resummation of planar diagrams is a complicated problem
in general, simplifications occur
in some cases, because supersymmetry 
leads to cancellations between various contributions.  
For certain quantities, the supersymmetry constraints are so strong
that all quantum corrections cancel. 
Such quantities do not depend on the coupling and can be
computed by summing tree-level diagrams. Non-renormalization theorems of this kind
are known to hold for
two and three point 
correlation functions
of chiral operators, which preserve 1/2 of 
\N supersymmetry \cite{Gub97}-\cite{Penati:1999ba}. 
The complete agreement of the supergravity predictions for these
correlators 
with field theory is one of the strongest
tests of the AdS/CFT correspondence \cite{D'Hoker:2002aw}.

Another quantity that is not affected by quantum corrections is an
expectation of the straight Wilson line. When properly defined, the
expectation value is equal to one, which is confirmed by perturbative computation
to the first two orders \cite{Erickson:2000af}, 
and also by semiclassical calculations in string theory \cite{Dru00'}.
A related quantity, the circular Wilson loop, can be obtained from the straight line
by a conformal transformation. Though SYM theory is conformally invariant,
the expectation value of the circular loop acquires
anomalous contributions \cite{Drukker:2000rr,Semenoff:2002kk} 
and is a non-trivial function of the
coupling constant. This function can be exactly calculated by summing 
Feynman diagrams that survive supersymmetry cancellations 
\cite{Erickson:2000af}.
Various correlation functions of the circular Wilson loop 
allow to trace how the sum of planar diagrams at weak coupling
transforms into the semiclassical string partition
function at strong coupling 
\cite{Drukker:2000rr,Erickson:2000af,Semenoff:2001xp,Semenoff:2002kk}.

The cancellation of quantum corrections for the straight Wilson line 
stems from supersymmetry. The Wilson line operator 
commutes with eight of the sixteen supercharges \cite{Drukker:1999zq}
and thus corresponds to a BPS state. 
The circular Wilson loop does not preserve any supersymmetry, but
commutes with eight linear combinations of
supersymmetry and superconformal generators \cite{Bianchi:2002gz}.
Simplifications that occur for the circular loop should be somehow related to its 
superconformal invariance.

I will construct Wilson loops that preserve 1/16, 1/8 or 1/4 of the
supersymmetry and study their expectation values 
 in perturbation theory and 
at strong coupling using AdS string theory. Comparing the results
of these calculations, I
conjecture that the
expectation values of 1/4 BPS Wilson loops are not renormalized. But less
supersymmetric Wilson loops do receive quantum corrections.

\newsection{Supersymmetry transformations}\label{2}

The field content of \N SYM theory consists of gauge fields $A_\mu$, six
scalars $\Phi_i$ ($i=4\ldots 9$) and four Majorana fermions $\Psi^A$,
all in the adjoint representation of $SU(N)$. It is convenient to
put fermions into a single Majorana-Weyl spinor of $Spin(9,1)$.
The Euclidean action then takes the following form:
\eq{
S=\frac{1}{g^2}\int d^4x\,\tr\left\{
\frac12\,F_{\mu\nu}^2+\br{D_\mu\Phi_i}^2-\frac12\,[\Phi_i,\Phi_j]^2
+\bar{\Psi}\Gamma^\mu D_\mu\Psi+i\bar{\Psi}\Gamma^i[\Phi_i,\Psi]
\right\},
}
where $\Gamma^M=(\Gamma^\mu,\Gamma^i)$ are ten-dimensional
Dirac matrices.
The supersymmetry transformations of the bosonic field are
\ar{\label{susy}
\D_\ep A_\mu&=&\bar{\Psi}\Gamma^\mu\ep, 
\non
\D_\ep \Phi_i&=&\bar{\Psi}\Gamma^i\ep,
}
where the parameter of transformation $\ep$ is a ten-dimensional Majorana-Weyl spinor.

The Wilson loop operator which is dual to the string in Anti-de-Sitter space
\cite{Maldacena:1998im}
is a hybrid of the usual non-Abelian phase factor and the scalar loop
of \cite{Makeenko:hm}:
\eq{\label{wdef}
W(C,\th)=\frac{1}{N}\,\tr{\rm P}\exp
\int ds\,\br{i A_\mu(x)\dot{x}^\mu+\Phi_i(x)\th^i|\dot{x}|}.
}
Here, $x^\mu(s)$ parameterizes the contour $C$ in $\mathbb{R}^4$
and $\th^i$ is a unit
six-vector: $\th^i\th^i=1$. This vector can depend on $s$,
though, in most of the papers on Wilson loops,
$\th^i$ was assumed to be constant. 

The supersymmetry variation of the Wilson loop is
\eq{
\D_\ep W(C,\th)=\frac{1}{N}\,\tr{\rm P}\int ds\,
\bar{\Psi}\br{i\Gamma^\mu \dx^\mu+\Gamma^i\th^i|\dx|}\ep
\,\exp
\int ds'\,\br{i A_\mu\dot{x}^\mu+\Phi_i\th^i|\dot{x}|}.
}
Some part of the
supersymmetry\footnote{It would be also interesting to consider
more general conditions for superconformal invariance of
a Wilson loop operator (see \cite{Bianchi:2002gz} for the discussion
of the circular loop).} 
will be preserved if
\eq{\label{bps}
\br{i\Gamma^\mu \dx^\mu+\Gamma^i\th^i|\dx|}\ep=0.
}
Since the linear combination of Dirac matrices 
$i\Gamma^\mu \dx^\mu+\Gamma^i\th^i|\dx|$ squares to zero,
equation \rf{bps} has eight independent solutions for any given 
$s$. In general, these solutions
 will depend on $s$, so an arbitrary Wilson loop
is only
locally supersymmetric. Local supersymmetry is not a symmetry 
of the action, however. The requirement that $\ep$ is $s$-independent
is a constraint on $x^\mu(s)$ and $\th^i(s)$. 
The number of linearly independent $\ep$'s that satisfy
eq.~\rf{bps}
determines the number of conserved supercharges.

It is easy to convince oneself that eq.~\rf{bps}
has no solutions for constant $\th^i$, unless $C$ is a straight line. 
Indeed, choosing parameterization of the contour $C$
such that $|\dx|=1$ and differentiating \rf{bps} in $s$, we get:
$$
i\Gamma^\mu \ddot{x}^\mu\,\ep=0,
$$
which implies that $\ddot{x}$ is identically zero.

For a general loop with varying $\th^i$ 
and a curved contour $C$, \rf{bps} constitutes an infinite set
of algebraic equations for sixteen unknown quantities.
In spite of the huge redundancy of these equations, they have non-trivial
solutions for certain $x^\mu$ and $\th^i$.
The complete classification of supersymmetric Wilson loops
is beyond the scope of the present paper. Instead, I will study a 
simple ansatz, for which \rf{bps} reduces to a finite number
of equations.  The ansatz amounts in requiring that the position of
the loop in $S^5$ 
follows the tangent vector $\dx^\mu$ of the
space-time contour $C$. The map from $S^3$ to $S^5$
is defined by an immersion of
$\mathbb{R}^4$ in $\mathbb{R}^6$ as a hyperplane:
\eq{
x^\mu\longmapsto x^\mu M_\mu^i,
}
where the rectangular matrix $M_\mu^i$ can be regarded
 as a projection operator:
\eq{\label{prj}
M_\mu^i M_\nu^i=\D_{\mu\nu}.
}
A particular form of $M_\mu^i$ is not important because of
$SO(4)\times SO(6)$ global symmetry of \N SYM.
Then,  
\eq{
\th^i=M^i_\mu \frac{\dx^\mu}{|\dx|}
}
maps a
tangent vector of the contour $C$ to a  
point on $S^5$. With this choice of $\th^i$, 
the Wilson loop operator becomes
\eq{\label{defsl}
W_s(C)=\frac{1}{N}\,\tr{\rm P}\exp\oint_C dx^\mu\,
\br{iA_\mu+M_\mu^i\Phi_i}.
}
This is my ansatz for 
the supersymmetric Wilson loop. The supersymmetry 
variation of this operator vanishes if
\eq{\label{susy0}
i\dx^\mu\br{\Gamma^\mu-iM_\mu^i\Gamma^i}\ep=0.
}
All $s$-dependence factors out,
 and one is left with four
algebraic equations:
\eq{\label{susy1}
\br{\Gamma^\mu-iM_\mu^i\Gamma^i}\ep=0.
}
These equations can be easily solved by choosing a particular basis
in the spinor representation of $Spin(10)$
\footnote{Strictly speaking, one should deal with $Spin(9,1)$
spinors, because of the Majorana condition on $\ep$. 
However, the signature of the
metric will not be important for the discussion below,
and I will
assume that the Dirac matrices anti-commute on $2\D^{MN}$
from now on, to simplify the notations.}.
Let us define four pairs of creation and annihilation operators:
\ar{
a^\mu&=&\frac{1}{2}\br{\Gamma^\mu-iM_\mu^i\Gamma^i}, 
\non
{a_\mu}\dd&=&\frac{1}{2}\br{\Gamma^\mu+iM_\mu^i\Gamma^i}.
}
The fifth pair of operators is constructed using two six-vectors
orthogonal to $M_\mu^i$:
\eq{
M_\mu^i v^i_{1,2}=0,~~~~~v^2_{1,2}=1,
}
\ar{
a^4&=&\frac{1}{2}\br{v_1^i\Gamma^i-iv_2^i\Gamma^i},
\non
{a_4}\dd&=&\frac{1}{2}\br{v_1^i\Gamma^i+iv_2^i\Gamma^i}.
}
The matrices $a^M$, ${a_M}\dd$ satisfy anti-commutation relations
\eq{
\{a^M,{a_N}\dd\}=\D^M_N,
}
and their Fock space can be identified with the
 spinor representation of $Spin(10)$. The chirality projection
leaves states with even (or odd, depending on the sign
of chirality projection) number of creation operators
acting on the Fock vacuum.

The equations \rf{susy1} in this representation are
\eq{
a^\mu|\ep\rangle=0,~~~~~\mu=0\ldots 3.
}
For the spinor to be annihilated by $a^0\ldots a^3$, the levels 
associated with these oscillators must be filled. 
There are two such states:
\ar{\label{sol}
|\ep_+\rangle&=&a_0\dd\ldots a_3\dd |0\rangle,
\non
|\ep_-\rangle&=&a_0\dd\ldots a_3\dd a_4\dd|0\rangle.
}
They have opposite chirality. 
Hence,
there is only one Weyl spinor that satisfies eqs.~\rf{susy1}
and, consequently, the Wilson loop operator
commutes with one of the sixteen supercharges.
So, a Wilson loop of the form \rf{defsl} preserves
supersymmetry and generically is 1/16 BPS.

\begin{table}[h]
\caption{\small The amount of supersymmetry 
for Wilson loops of various dimensions.
}
\label{tab1}
\begin{center}
\begin{tabular}{|c|c|}
\hline
Dimensionality of the loop & Amount of supersymmetry \\
\hline
4D & 1/16 \\
3D & 1/8 \\
2D & 1/4 \\
1D & 1/2 \\
\hline
\end{tabular}
\end{center}
\end{table}

 The supersymmetry
is enhanced if the contour $C$ has a special shape.
Consider, for instance, a spatial Wilson loop which
lies in a three-dimensional time slice $x^0=0$.
$\dx^0$ is identically equal to zero in this case, so only three of four
constraints in \rf{susy1}, those with $\mu=1\ldots 3$, should be 
imposed to satisfy \rf{susy0}. Then $|\ep\rangle$ must be
annihilated only by three oscillators. In addition to \rf{sol},
there are two extra solutions that satisfy these constraints:
\ar{\label{sol+}
|\ep_+^{(1)}\rangle&=&a_0\dd\ldots a_3\dd |0\rangle,
\non
|\ep_-^{(1)}\rangle&=&a_0\dd\ldots a_3\dd a_4\dd|0\rangle,
\non
|\ep_+^{(2)}\rangle&=&a_1\dd\ldots a_3\dd |0\rangle,
\non
|\ep_-^{(2)}\rangle&=&a_1\dd\ldots a_3\dd a_4\dd|0\rangle.
}
Two of these spinors are chiral, so 
a three-dimensional loop commutes with 2 supercharges and
preserves 1/8 of the
supersymmetry. If the contour $C$ lies in a two-dimensional plane,
the number of supersymmetries again doubles,
 and the planar Wilson loop preserves 1/4 of the supersymmetry.
A one-dimensional supersymmetric Wilson loop is a familiar
Wilson line with constant $\th^i$, which is 1/2 BPS.
The amount of supersymmetry as a function of 
dimensionality of the loop is summarized in table~\ref{tab1}.

\newsection{Supersymmetric Wilson loops in perturbation theory}

One may anticipate a lot of cancellations 
between quantum corrections for the 
supersymmetric Wilson loops.
I will calculate the expectation value 
\eq{
\vev{W_s(C)}=\vev{\frac{1}{N}\,\tr{\rm P}\exp\oint_C dx^\mu\,
\br{iA_\mu+M_\mu^i\Phi^i}}
}
to the two first orders in perturbation theory and
to the leading order in the large-$N$ expansion and will
demonstrate that all corrections mutually cancel.

\begin{figure}[h]
\begin{center}
\epsfxsize=9cm
\epsfbox{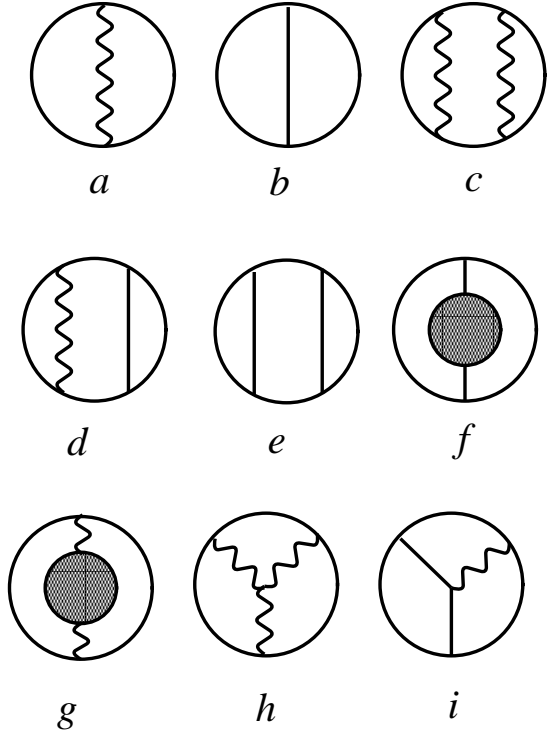}
\end{center}
\caption[x]{\small Feynman diagrams of leading and next-to-leading 
orders.}
\label{diagr}
\end{figure}

There are nine diagrams  (fig.~\ref{diagr}) that contribute
at $O((g^2N)^2)$.
Each of the individual graphs in fig.~\ref{diagr}
contains UV divergences. 
The divergences cancel in the sum, but intermediate calculations
require a regularization. An explicit regularization prescription
will not be important for the calculations below, as soon
as the regularization preserves supersymmetry. One can always keep in mind
the dimensional reduction which works 
to this order in perturbation theory 
\cite{Erickson:2000af,Plefka:2001bu,Arutyunov:2001hs}.
I will not even need an explicit form of
propagators. The only necessary ingredient is the equality of
gauge boson and scalar propagators in the Feynman
gauge, which is a consequence of supersymmetry:
\ar{
\vev{A_\mu^{ab}(x)A_\nu^{cd}(y)}_0
&=&g^2\br{\D^{ad}\D^{bc}-\frac{1}{N}\,\D^{ab}\D^{cd}}
\D_{\mu\nu}D(x-y),
\non
\vev{\Phi_i^{ab}(x)\Phi_i^{cd}(y)}_0
&=&g^2\br{\D^{ad}\D^{bc}-\frac{1}{N}\,\D^{ab}\D^{cd}}
\D_{ij}D(x-y),
} 
where $a,b,c,d$ are $SU(N)$ indices. 

The diagrams of the leading order, (a) and (b), mutually
cancel, because
\eq{
\vev{\br{iA_\mu(x)+M_\mu^i\Phi_i(x)}
\br{iA_\nu(y)+M_\nu^j\Phi_j(y)}}_0
\propto
(i^2\D_{\mu\nu}+M_\mu^iM_\nu^i)D(x-y)=0,
}
owing to the identity \rf{prj}. All diagrams without
internal vertices cancel for the same reason,
in particular, the diagrams
 (c), (d) and (e). 
As was shown in \cite{Erickson:2000af}, one-loop
corrected scalar and vector propagators are still equal,
up to total derivatives,
so the diagrams (f) and (g) also sum up to zero. 

The remaining graphs (h) and (i) are computed as follows:
\ar{
(h)+(i)&=&
\frac{2}{g^2N}\int d^4x\,\oint dx_1^\mu dx_2^\nu dx_3^\lambda\,
\th_c(x_1,x_2,x_3)
\non
&&\times
\left(\frac13\,\vev{
\tr A_\mu(x_1)A_\nu(x_2)A_\lambda(x_3)\,
\tr \d_\rho A_\sigma(x)[A_\rho(x),A_\sigma(x)]}
\right.
\non
&&-
\left.\vphantom{\frac13}
\vev{M^i_\mu M^j_\nu\tr\Phi_i(x_1)\Phi_j(x_2)A_\lambda(x_3)\,
\tr \d_\rho\Phi_k(x)[A_\rho(x),\Phi_k(x)]
}_0\right)
\non
&=&
2(g^2N)^2\int d^4x\,\oint dx_1^\mu dx_2^\nu dx_3^\lambda\,
\th_c(x_1,x_2,x_3)
\D_{\mu\nu}D(x-x_3)
\non
&&\times 
\Bigl(D(x-x_1)\d_\lambda D(x-x_2)
-\d_\lambda D(x-x_1)D(x-x_2)
\non
&&
-D(x-x_1)\d_\lambda D(x-x_2)
+\d_\lambda D(x-x_1)D(x-x_2)\Bigr)
=0.
}
Here, $\th_c(x_1,x_2,x_3)$ is equal to one, if points $x_1$,
$x_2$ and $x_3$ are cyclically ordered along the contour $C$,
 and is equal to zero otherwise.

All corrections of orders $g^2N$ and $(g^2N)^2$ cancel.
That may indicate that expectation values of supersymmetric
Wilson loops obey non-renormalization theorems. 
If so,  AdS/CFT calculations at strong coupling must
also give $\vev{W_s(C)}=1$.  

\newsection{Supersymmetric Wilson loops in string theory}

According to the AdS/CFT conjecture, Wilson loops couple
directly to strings that propagate in $AdS_5\times S^5$.
The expectation value of the Wilson loop is the string
partition function \cite{Maldacena:1998im,Rey:1998ik}:
\ar{\label{Z}
\vev{W(C,\th)}&=&
\int_{\rm reg} DX^M D\vartheta^\alpha Dh_{ab}\,
\exp\left(-\frac{\st}{4\pi}\,\int_D d^2\sigma\,\sqrt{h}h^{ab}G_{MN}
\d_aX^M\d_bX^N
\right.
\non
&&\left.\vphantom{\int_{\rm reg} DX^M D\vartheta^\alpha Dh_{ab}\,
\exp-\frac{\st}{4\pi}\,\int_D d^2\sigma\,\sqrt{h}h^{ab}G_{MN}
\d_aX^M\d_bX^N}
+{\rm fermions}\right),
}
The sigma-model metric is defined by the line element:
\eq{
\ds^2=\frac{1}{Z^2}(dX_\mu^2+dZ^2)+d\Omega_{S^5}^2.
}
The fermionic part of the action is known
\cite{Metsaev:1998it}-\cite{Frolov:2002av},
\cite{Dru00'}, but will not be used 
here. The string world sheet extends to the boundary of AdS space, where 
it terminates on the contour $C$. In other words, 
the sigma-model path integral is
supplemented by the boundary conditions:
\eq{
\bc{X^\mu}=x^\mu(s),~~~~~\bc{Z}=\es,~~~~~\bc{\Theta^i}=\th^i(s).
}
$\Theta^i$ parameterizes a position of the world sheet in $S^5$.
The regularization parameter $\es$ cuts off  divergences which 
arise because of the $1/z^2$ singularity of the metric at the boundary.
At the end, $\es$ should be sent to zero. If the string partition function
is appropriately defined, the divergences
appear only in the intermediate calculations and eventually cancel.
The correct definition of the partition function involves the Legendre
transform in $Z$ \cite{Drukker:1999zq}. 
Explicit implementation of the Legendre transform
is somewhat cumbersome, but, fortunately, 
in the semiclassical approximation, the Legendre
transform amounts in dropping $1/\es$ divergences whenever they appear.

It is not known how to solve the $AdS_5\times S^5$ sigma model
exactly. The only simplification occurs at large 't~Hooft coupling,
when the sigma model becomes 
weakly coupled,
and
the partition functions can be computed in the saddle-point
 approximation. Minima of the string action correspond 
to minimal
surfaces in $AdS_5\times S^5$, whose boundary is
the contour $C$. The action at the saddle point
is the  area of the minimal surface:
\eq{
A(C)=\int d^2\sigma\,\sqrt{\det_{ab}G_{MN}\d_a\cl^M\d_b\cl^N},
}
which should be regularized by subtraction of the boundary divergence.
The $\alpha'$ expansion of the sigma model, obtained by
expanding around the classical solution and integrating out fluctuations,
yields $1/\st$ expansion of the Wilson loop expectation value.
There are additional complications if
the classical solution depends on parameters (has moduli). 
The parameters give rise
to zero mode fluctuations which should be separated by introducing 
collective coordinates for the moduli of the classical solution. 
Each moduli integration is accompanied by a factor of
$\alpha'{}^{1/2}$, which, in the present case, should be identified
with $(g^2N)^{1/4}$. The gauge fixing of the world sheet diffeomorphism /
Weyl invariance also produces a non-trivial factor, because the usual
conformal gauge leaves residual three-parametric gauge freedom in the disk
partition function \cite{Alvarez:1982zi}. The
residual gauge symmetries give a factor of $(g^2N)^{-1/4}$ each
\cite{Drukker:2000rr}.
Consequently, 
the semiclassical partition function for the Wilson loop vev
has the following general form:
\eq{
\vev{W(C,\th)}=\const\,(g^2N)^{(N_{z.m.}-3)/4}
\exp\br{-\frac{\st}{2\pi}\,A(C)} + 1/\st{\rm~correction},
}
where $N_{z.m.}$ is the number of zero modes, or the number of
moduli in the classical solution $\cl^M$. An overall constant
comes from the integration over non-zero-mode fluctuations.

Potential non-renormalization of supersymmetric Wilson loops implies that 
corresponding minimal surfaces have zero area.
For loops with constant $\th_i$, which have
 been most extensively studied so far, 
the minimal surface sits at
one point in $S^5$ and extends only in $AdS_5$. It can be shown that
the regularized area of a minimal surface in $AdS_5$ 
is strictly negative because of the subtraction of 
the boundary divergence.
 The area in $S^5$ is positive,
which may well cancel the $AdS_5$ contribution. So, nullification 
of the classical string action is not very surprising, and I will 
show that it indeed happens for the simplest contours.
The cancellation of the zero mode factor is a much more unexpected
fact. For the Wilson loop vev to be one, the minimal surface must be 
degenerate and must depend exactly on three parameters.
A minimal surface with a given boundary in $AdS_5$ is usually unique.
 Several degenerate minima
of the string action can coexist in some special cases, but
this degeneracy is never parametric. This leads to a characteristic
$(g^2N)^{-3/4}$ pre-factor in the Wilson loop expectation value
\cite{Drukker:2000rr}
which is confirmed by exact field-theory calculations 
\cite{Erickson:2000af}.
It turns out that a non-trivial dependence on $S^5$ coordinates
qualitatively changes the situation. Minimal surfaces which are
associated with supersymmetric Wilson loops are always
parametrically degenerate and have moduli. The number of
moduli turns out to depend on the amount of supersymmetry
preserved by the Wilson loop. 
Minimal surfaces with constant position
in $S^5$ can be regarded as a degenerate case, in which the moduli 
space shrinks to zero size.

The simplest  supersymmetric Wilson loop
consists of two anti-parallel lines with points at
infinity identified, which can be thought of as a limit
of an infinitely long rectangular contour. Supersymmetry requires that $\th_i$
follows the tangent vector of the contour $C$. Since the tangent
vector rotates through $\pi$ at infinity,  
each of the two lines should be put at diametrically opposite points 
on $S^5$. This is a particular case of configuration considered in 
\cite{Maldacena:1998im}. 
It was observed there that the potential between anti-parallel
lines vanishes when their angular separation on $S^5$
reaches $\pi$. This is exactly where the Wilson loop becomes supersymmetric.

I will consider in detail another simple example, the circular
supersymmetric Wilson loop. Due to the scale invariance of
\N SYM, the radius of the circle can be put to one, and it
can be parameterized as
\eq{
x^\mu(s)=(\cos s,\sin s,0,0).
}
The loop in $S^5$ must be an equatorial circle to satisfy
the supersymmetry constraints.
Choosing the angular parameterization of $S^5$, when
the metric is
\eq{\label{1+3}
d\Omega^2_{S^5}=d\psi^2+\cos^2\psi\,d\ph^2+\sin^2\psi\,
d\Omega^2_{S^3},~~~~~0\leq \psi<\pi,~~~~~0\leq\ph<2\pi,
}
we have the following boundary conditions for the $S^5$ coordinates
\eq{\label{ssl}
\ph(s)=s,~~~~~\psi(s)=0,~~~~~{\bf n}=\const,
}
where ${\bf n}$ is a unit four-vector that corresponds to a point
on $S^3$.
The minimal surface can be parameterized by the angle $s$ and 
the AdS scale $z$. Imposing the symmetry constraints, we can
look for a solution of the following form:
\eq{
X^\mu=(R\cos s,R\sin s,0,0),~R\equiv R(z),~~~~~Z=z,~~~~~\varPhi=s,
~~~~~\varPsi\equiv\varPsi(z).
}
The area of this surface is
\eq{
A=2\pi\int dz\,\sqrt{\br{\frac{R^2}{z^2}+\cos^2\varPsi}
\br{\varPsi'{}^2+\frac{R'{}^2+1}{z^2}}}.
}
Variational equations following from minimization of $A$
are quite complicated, but the solution is rather simple:
\eq{\label{zg}
R=\sqrt{1-z^2},~~~~~\sin\varPsi=z.
}
Checking that this solves equations of motion
for the Nambu-Goto action
is a straightforward but lengthy exercise. An easier way to 
see that this corresponds to a minimal surface is outlined below.

A metric 
induced on the minimal 
surface is
\eq{
\ds^2=\br{\frac{1}{z^2}+1}\left[\frac{dz^2}{1-z^2}+(1-z^2)ds^2
\right],
}
and the minimal area is readily computed
\eq{
A(\cir)=\int_0^{2\pi}ds\int_\es^1dz\,\br{\frac{1}{z^2}+1}=\frac{2\pi}{\es}.
}
The regularized area, left after subtraction of the divergence,
is zero, which is consistent with non-renormalization
of the Wilson loop expectation value: $\vev{W_s(\cir)}=1$.

The above solution can be easily obtained if we start with the action in 
the Polyakov form, as in \rf{Z}, and fix the conformal gauge:
$h_{ab}=\D_{ab}$. The equations for the minimal surface then
follow from the action
\eq{
A=\pi\int d\tau\,\left[\frac{R'{}^2+R^2+Z'{}^2}{Z^2}
+\varPsi'{}^2+\cos^2\varPsi\right],
}
where now $R$, $Z$ and $\varPsi$ are functions of $\tau$.
The equations should be supplemented by Virasoro
constraints,
which require the induced metric to be unit matrix up to
a conformal factor:
\eq{
G_{MN}\d_a X^M\d_b X^N=\e^\phi\D_{ab}.
}
Both the equations of motion and the Virasoro constraints
can be solved separately for the $AdS_5$ and the $S^5$
coordinates of the string world sheet. Hence, the minimal surface in 
the conformal gauge is a direct sum of two surfaces
in $AdS_5$ and in $S^5$. Minimal surface in $AdS_5$ that 
has a circle as its boundary is known 
\cite{Drukker:1999zq,Berenstein:1999ij}. 
Transforming the solution of
\cite{Drukker:1999zq,Berenstein:1999ij} 
to the conformal gauge we get:
\eq{\label{adscg}
R=\frac{1}{\cosh \tau},~~~~~Z=\tanh\tau.
}
The $AdS_5$ part of the induced metric is
\eq{
\ds^2_{AdS_5{\rm ,induced}}=\frac{1}{\sinh^2\tau}(d\tau^2+ds^2).
}
Solving the equation of motion for $\varPsi$ with boundary conditions
\rf{ssl}, we find:
\eq{\label{scg}
\sin\varPsi=\tanh\tau,
}
and
\eq{
\ds^2_{S^5{\rm,induced}}
=\frac{1}{\cosh^2\tau}(d\tau^2+ds^2).
}
Changing the world-sheet coordinates from $(\tau, s)$
to $(z,s)$ in \rf{adscg}, \rf{scg} we get back to \rf{zg}.
Computation of the area in the conformal gauge, of course,
gives the same result:
\eq{
A(\cir)=2\pi\int_\es^\infty d\tau\,\br{\frac{1}{\sinh^2\tau}
+\frac{1}{\cosh^2\tau}}=
2\pi\left.(\tanh\tau-\coth\tau)\right|_\es^\infty
=\frac{2\pi}{\es}\rightarrow 0.
}

The above solution for the minimal surface depends on an arbitrary
point on $S^3$, denoted by ${\bf n}$ in \rf{ssl}. Different ${\bf n}$'s
correspond to different minimal surfaces, so the solution indeed depends on
three arbitrary parameters, exactly as predicted by supersymmetry!
Why minimal surfaces that sit at one point in $S^5$ do not have this
degeneracy? The reason is that  $S^3$ shrinks to zero size at $\psi=0$.
 $(\psi=0,{\bf n})$ correspond to one point in $S^5$ 
for all ${\bf n}$. As soon as
$\varPsi$ is identically equal to zero, the degeneracy does not arise.
It is easy to see that the three-parametric degeneracy is not
specific to the circular loop. Indeed, any planar contour projects
onto an equatorial circle in $S^5$ under the map defined in sec.~\ref{2}.
For symmetry reasons, the minimal surface will not extend
into the orthogonal $S^3$, coordinates on which can be regarded as
moduli parameterizing the minimal surface.

It is natural to assume that $\vev{W_s(C)}=1$ for any planar loop $C$. 
This conjecture relies on the nullification of the minimal area 
in two cases, for
the anti-parallel lines \cite{Maldacena:1998im} and for
 the circular loop considered above.
It would clearly 
be desirable to have a general proof 
that the minimal area is zero for any planar
contour. 
On the other hand, cancellation of zero modes simply
follows from the '3+1' decomposition of the metric in
eq.~\rf{1+3} and applies to any 1/4 BPS Wilson loop.

What about less supersymmetric loops? A $d$-dimensional
contour in $\mathbb{R}^4$ maps onto a curve in the equatorial
$S^{d-1}$ in $S^5$.
It is then convenient to decompose the $S^5$ metric
in the '($d$-1)+(5-$d$)' way:
\eq{
d\Omega_{S^5}^2=d\psi^2+\cos^2\psi\,d\Omega_{S^{d-1}}^2
+\sin^2\psi\,d\Omega_{S^{5-d}}^2.
}
The minimal surface will sit at one point in $S^{5-d}$.
So, the solution for the minimal surface will contain
$N_{z.m.}=(5-d)$ moduli.
Consequently, the expectation value for a $d$-dimensional
supersymmetric Wilson loop is proportional 
to\footnote{This formula is not applicable at $d=1$ (the straight line),
because of boundary conditions at spatial infinity that
should be imposed in this case. These extra boundary conditions
modify zero mode counting. 
Properly defined partition function
for the straight Wilson line should be equal to one.}
\eq{\label{ddim}
\vev{W_s(C_d)}\propto (g^2N)^{(2-d)/4}\exp\br{-\frac{\st}{2\pi}\,A(C)}.
}
If $d>2$, the expectation value is a non-trivial function of
the 't~Hooft coupling, whatever the minimal area is. There are no 
reasons to expect that 
the minimal area for non-planar contours is zero, since
non-renormalization does not work in this case anyway.

\newsection{Discussion}

The supersymmetric Wilson loops can be constructed in ${\cal N}=2$
SYM theory as well. The details are given in  Appendix. 
These operators can be useful in
the study of supergravity duals of ${\cal N}=2$ gauge theories
constructed in \cite{Bertolini:2000dk}-\cite{Bertolini:2001gq}.

It is likely that planar supersymmetric Wilson loops in \N SYM,
which preserve 1/4 of the supersymmetry, are not renormalized by 
quantum corrections. At strong coupling, non-renormalization
of planar Wilson loops implies cancellation of the minimal area
in $AdS_5\times S^5$ for rather wide class of boundary conditions.
This is a  very non-trivial statement about minimal
surfaces in $AdS_5\times S^5$. I did not prove it, but checked for
the circular and the rectangular loops. 

Wilson loops with lesser degree of supersymmetry do renormalize.
A non-trivial dependence on 't~Hooft coupling
follows from zero mode counting in the string partition 
function. The dependence of non-planar loops on 't~Hooft
coupling and the coupling-constant independence of planar loops 
has an interesting consequence. Consider a
slightly non-planar Wilson loop; let $\D$ be a parameter of
non-planarity. The strong-coupling asymptotics of the
Wilson loop vev then is a discontinuous function of $\D$.
This can be seen from the structure
 of zero-mode contribution (a pre-factor 
in \rf{ddim}) to the string partition function. The number of zero modes
changes as $\D$ turns to zero, so the expectation values for $\D=0$ and 
for infinitesimally small but non-zero $\D$  differ by a finite amount. Such a 
discontinuity is absent in any order of perturbation theory, as any Feynman
diagram is an analytic function of $\D$. Similar strong-coupling
'phase transitions' occur in Wilson loop correlators
\cite{Gross:1998gk}-\cite{Zarembo:2001jp}, 
they arise due to the string breaking. 

\subsection*{Acknowledgements}

I am grateful to Anton Alekseev for hospitality at the University of
Geneva, where this work was completed. The work was supported
by STINT grant IG 2001-062 and by Royal Swedish Academy of Sciences
and, in part, by RFBR grant 02-02-17260 and grant
00-15-96557 for the promotion of scientific schools.

\setcounter{section}{0}
\appendix{Supersymmetric Wilson loops in ${\cal N}=2$ SYM theory}

The field content of ${\cal N}=2$ SYM theory consists of the gauge fields
$A_\mu$, two Majorana fermions $\lambda^A_\alpha$, $A=1,2$, and a
complex scalar $\Phi$. All fields are in the adjoint representation of
the gauge group, $SU(N)$. The supersymmetry
transformations of the bosonic fields are:
\ar{
\D A_\mu&=&-i\bar{\lambda}_A\bs_\mu\ep^A+i\bar{\ep}_A\bs_\mu\lambda^A,
\non
\D \Phi&=&\sqrt{2}\,\es_{AB}\ep^A\lambda^B,
\non
\D \Phi\dd&=&-\sqrt{2}\,\es^{AB}\bar{\ep}_A\bar{\lambda}_B,
}
where $\ep^A$ are two Majorana spinors, and
I follow the conventions of \cite{Wess}. In particular,
I switch to 
 the Minkowski metric
with signature $(-+++)$.

The supersymmetric Wilson loop is defined as
\eq{\label{n2}
W_s(C,\zeta)=\frac{1}{N}\,\tr{\rm P}\exp\int ds\,
\br{iA_\mu\dx^\mu+\Phi\zeta^*|\dx|+\Phi^*\zeta|\dx|},
}
where $\zeta\equiv\zeta(s)$ is an arbitrary complex-valued 
function of the parameter on the
contour, which satisfies $|\zeta|=1/\sqrt{2}$.
 The modulus of the tangent vector, $|\dx|$, 
stands for $\sqrt{\dx^2}$ and is imaginary for 
time-like contours. Therefore,  time-like Wilson loops are
associated with unitary matrices, as usual, and it is only space-like loops that contain
a Hermitian piece in the exponent. Light-like loops do not couple to scalars at all.

The requirement that the supersymmetry variation of the Wilson loop turns to zero
is equivalent to the following equations:
\ar{
\bs^\mu \dx_\mu\ep^A&=&-\sqrt{2}\,|\dx|\zeta\es^{AB}\bar{\ep}_B,
\non
\bar{\ep}_A\bs^\mu\dx_\mu&=&\sqrt{2}\,|\dx|\zeta^*\ep^B\es_{BA}.
}
These equations are Hermitian conjugates of one another.
The first equation can be used to express $\bar{\ep}_A$ 
in terms of $\ep^A$:
\eq{\label{onlyc}
\bar{\ep}_A=-\frac{1}{\sqrt{2}\zeta |\dx|}\,
\es_{AB}\bs^\mu \dx_\mu\ep^B.
}
Substituting this into the second equation we get
$$
\bs^{\mu\dot{\alpha}\alpha}\es_{\dot{\alpha}\dot{\beta}}
\bs^{\nu\dot{\beta}\beta}\dx_\mu\dx_\nu\es_{AB}\ep^B_\alpha
=-2|\zeta|^2\dx^2\es_{AB}\ep^{B\beta},
$$
which can be shown to be an identity, provided $2|\zeta|^2=1$.

The four constraints \rf{onlyc} fix half of the parameters
of the supersymmetry transformation. Again, generic Wilson loop preserves
only local supersymmetry, since the right hand side
of \rf{onlyc} in general depends on a position on the curve.
An obvious exception is the straight line with constant $\zeta$, 
which preserves 1/2 of global ${\cal N}=2$
supersymmetry. It would be interesting to understand if there are 
other BPS Wilson loops, which preserve smaller part
of the supersymmetry.


\begin{thebibliography}{99}

\bibitem{Maldacena:1998re}
J.~Maldacena,
``The large N limit of super-conformal field theories and supergravity,''
Adv.\ Theor.\ Math.\ Phys.\  {\bf 2}, 231 (1998)
[Int.\ J.\ Theor.\ Phys.\  {\bf 38}, 1113 (1998)]
[hep-th/9711200].

\bibitem{Gubser:1998bc}
S.~S.~Gubser, I.~R.~Klebanov and A.~M.~Polyakov,
``Gauge theory correlators from non-critical string theory,''
Phys.\ Lett.\ B {\bf 428} (1998) 105
[arXiv:hep-th/9802109].

\bibitem{Witten:1998qj}
E.~Witten,
``Anti-de Sitter space and holography,''
Adv.\ Theor.\ Math.\ Phys.\  {\bf 2}, 253 (1998)
[hep-th/9802150].

\bibitem{Aharony:1999ti}
O.~Aharony, S.~S.~Gubser, J.~Maldacena, H.~Ooguri and Y.~Oz,
``Large N field theories, string theory and gravity,''
Phys.\ Rept.\  {\bf 323} (2000) 183
[arXiv:hep-th/9905111].

\bibitem{'tHooft:1974jz}
G.~'t Hooft,
``A Planar Diagram Theory For Strong Interactions,''
Nucl.\ Phys.\ B {\bf 72}, 461 (1974).

\bibitem{Gub97}
S.S.~Gubser and I.R.~Klebanov,
``Absorption by branes and Schwinger terms in the world volume theory,"
Phys. Lett. {\bf B413}, 41 (1997)
[arXiv:hep-th/9708005].

\bibitem{Ans97}
D.~Anselmi, D.Z.~Freedman, M.T.~Grisaru and A.A.~Johansen,
``Nonperturbative formulas for central functions of supersymmetric gauge
theories,"
Nucl. Phys. {\bf B526}, 543 (1998)
[arXiv:hep-th/9708042].

\bibitem{Fre98}
D.Z.~Freedman, S..D.~Mathur, A.~Matusis and L.~Rastelli,
``Correlation functions in the CFT(d) / AdS(d+1) correspondence,"
Nucl. Phys. {\bf B546}, 96 (1999)
[arXiv:hep-th/9804058].

\bibitem{Lee98}
S.~Lee, S.~Minwalla, M.~Rangamani and N.~Seiberg,
``Three point functions of chiral operators in D = 4, N=4 SYM at large N,"
Adv. Theor. Math. Phys. {\bf 2}, 697 (1998)
[arXiv:hep-th/9806074].

\bibitem{DHo98}
E.~D'Hoker, D.Z.~Freedman and W.~Skiba,
``Field theory tests for correlators in the AdS / CFT correspondence,"
Phys. Rev. {\bf D59}, 045008 (1999)
[arXiv:hep-th/9807098].

\bibitem{Intriligator:1998ig}
K.~A.~Intriligator,
``Bonus symmetries of N = 4 super-Yang-Mills correlation functions via  AdS duality,''
Nucl.\ Phys.\ B {\bf 551} (1999) 575
[arXiv:hep-th/9811047].

\bibitem{Gonzalez-Rey:1999ih}
F.~Gonzalez-Rey, B.~Kulik and I.~Y.~Park,
``Non-renormalization of two point and 
three point correlators of N = 4  SYM in N = 1 superspace,''
Phys.\ Lett.\ B {\bf 455} (1999) 164
[arXiv:hep-th/9903094].

\bibitem{Intriligator:1999ff}
K.~A.~Intriligator and W.~Skiba,
``Bonus symmetry and the operator product expansion of N = 4  super-Yang-Mills,''
Nucl.\ Phys.\ B {\bf 559} (1999) 165
[arXiv:hep-th/9905020].

\bibitem{Eden:1999gh}
B.~Eden, P.~S.~Howe and P.~C.~West,
``Nilpotent invariants in N = 4 SYM,''
Phys.\ Lett.\ B {\bf 463} (1999) 19
[arXiv:hep-th/9905085].

\bibitem{Petkou:1999fv}
A.~Petkou and K.~Skenderis,
``A non-renormalization theorem for conformal anomalies,''
Nucl.\ Phys.\ B {\bf 561} (1999) 100
[arXiv:hep-th/9906030].

\bibitem{Liu:1999kg}
H.~Liu and A.~A.~Tseytlin,
``Dilaton-fixed scalar correlators and AdS(5) x S(5) - SYM  correspondence,''
JHEP {\bf 9910} (1999) 003
[arXiv:hep-th/9906151].

\bibitem{Penati:1999ba}
S.~Penati, A.~Santambrogio and D.~Zanon,
``Two-point functions of chiral operators in N = 4 SYM at order $g^4$,''
JHEP {\bf 9912} (1999) 006
[arXiv:hep-th/9910197];
``More on correlators and contact terms in N = 4 SYM at order $g^4$,''
Nucl.\ Phys.\ B {\bf 593} (2001) 651
[arXiv:hep-th/0005223].

\bibitem{D'Hoker:2002aw}
E.~D'Hoker and D.~Z.~Freedman,
``Supersymmetric gauge theories and the AdS/CFT correspondence,''
arXiv:hep-th/0201253.

\bibitem{Erickson:2000af} J.~K.~Erickson, 
G.~W.~Semenoff and K.~Zarembo, 
``Wilson loops in N = 4 supersymmetric 
Yang-Mills theory,'' 
Nucl.\ Phys.\ B {\bf 582}, 155 (2000) 
[arXiv:hep-th/0003055]. 

\bibitem{Dru00'}
N.~Drukker, D.~J.~Gross and A.~Tseytlin,
``Green-Schwarz string in AdS(5) x S(5): 
Semiclassical partition  function,''
JHEP{\bf 0004}, 021 (2000)
[arXiv:hep-th/0001204].

\bibitem{Drukker:2000rr}
N.~Drukker and D.~J.~Gross,
``An exact prediction of N = 4 SUSYM theory for string theory,''
J.\ Math.\ Phys.\  {\bf 42} (2001) 2896
[arXiv:hep-th/0010274].

\bibitem{Semenoff:2002kk}
G.~W.~Semenoff and K.~Zarembo,
``Wilson loops in SYM theory: From weak to strong coupling,''
Nucl.\ Phys.\ Proc.\ Suppl.\  {\bf 108} (2002) 106
[short version of arXiv:hep-th/0202156].

\bibitem{Semenoff:2001xp}
G.~W.~Semenoff and K.~Zarembo,
``More exact predictions of SUSYM for string theory,''
Nucl.\ Phys.\ B {\bf 616} (2001) 34
[arXiv:hep-th/0106015].

\bibitem{Drukker:1999zq}
N.~Drukker, D.~J.~Gross and H.~Ooguri,
``Wilson loops and minimal surfaces,''
Phys.\ Rev.\ D {\bf 60}, 125006 (1999)
[arXiv:hep-th/9904191].

\bibitem{Bianchi:2002gz}
M.~Bianchi, M.~B.~Green and S.~Kovacs,
``Instanton corrections to circular Wilson loops in N = 4 supersymmetric  Yang-Mills,''
JHEP {\bf 0204} (2002) 040
[arXiv:hep-th/0202003].

\bibitem{Maldacena:1998im}
J.~Maldacena,
``Wilson loops in large N field theories,''
Phys.\ Rev.\ Lett.\  {\bf 80}, 4859 (1998)
[arXiv:hep-th/9803002].

\bibitem{Makeenko:hm}
Y.~M.~Makeenko,
``Polygon Discretization Of The Loop Space Equation,''
Phys.\ Lett.\ B {\bf 212} (1988) 221.

\bibitem{Plefka:2001bu}
J.~Plefka and M.~Staudacher,
``Two loops to two loops in N = 4 supersymmetric Yang-Mills theory,''
JHEP {\bf 0109}, 031 (2001)
[arXiv:hep-th/0108182].

\bibitem{Arutyunov:2001hs}
G.~Arutyunov, J.~Plefka and M.~Staudacher,
``Limiting geometries of two circular Maldacena-Wilson loop operators,''
JHEP {\bf 0112} (2001) 014
[arXiv:hep-th/0111290].

\bibitem{Rey:1998ik}
S.~J.~Rey and J.~Yee,
``Macroscopic strings as heavy quarks in large N gauge theory and  anti-de Sitter supergravity,''
Eur.\ Phys.\ J.\ C {\bf 22} (2001) 379
[arXiv:hep-th/9803001].

\bibitem{Metsaev:1998it}
R.~R.~Metsaev and A.~A.~Tseytlin,
``Type IIB superstring action in AdS(5) x S(5) background,''
Nucl.\ Phys.\ B {\bf 533} (1998) 109
[arXiv:hep-th/9805028].

\bibitem{Kallosh:1998zx}
R.~Kallosh, J.~Rahmfeld and A.~Rajaraman,
``Near horizon superspace,''
JHEP {\bf 9809} (1998) 002
[arXiv:hep-th/9805217].

\bibitem{Pesando:1998fv}
I.~Pesando,
``A kappa gauge fixed type IIB superstring action on AdS(5) x S(5),''
JHEP {\bf 9811} (1998) 002
[arXiv:hep-th/9808020].

\bibitem{Kallosh:1998nx}
R.~Kallosh and J.~Rahmfeld,
``The GS string action on AdS(5) x S(5),''
Phys.\ Lett.\ B {\bf 443} (1998) 143
[arXiv:hep-th/9808038].

\bibitem{Kallosh:1998ji}
R.~Kallosh and A.~A.~Tseytlin,
``Simplifying superstring action on AdS(5) x S(5),''
JHEP {\bf 9810} (1998) 016
[arXiv:hep-th/9808088].

\bibitem{Frolov:2002av}
S.~Frolov and A.~A.~Tseytlin,
``Semiclassical quantization of rotating superstring in $AdS_5\times S^5$,''
JHEP {\bf 0206} (2002) 007
[arXiv:hep-th/0204226].

\bibitem{Alvarez:1982zi}
O.~Alvarez,
``Theory Of Strings With Boundaries: Fluctuations, Topology, And Quantum Geometry,''
Nucl.\ Phys.\ B {\bf 216} (1983) 125.
 
\bibitem{Berenstein:1999ij} D.~Berenstein, R.~Corrado, 
W.~Fischler and J.~Maldacena, 
``The operator product expansion for Wilson loops and surfaces in the
large N limit,'' Phys.\ Rev.\ D {\bf 59}, 105023 (1999)
[arXiv:hep-th/9809188].

\bibitem{Bertolini:2000dk}
M.~Bertolini, P.~Di Vecchia, M.~Frau, A.~Lerda, R.~Marotta and I.~Pesando,
``Fractional D-branes and their gauge duals,''
JHEP {\bf 0102} (2001) 014
[arXiv:hep-th/0011077].

\bibitem{Polchinski:2000mx}
J.~Polchinski,
``N = 2 gauge-gravity duals,''
Int.\ J.\ Mod.\ Phys.\ A {\bf 16} (2001) 707
[arXiv:hep-th/0011193].

\bibitem{Bertolini:2001qa}
M.~Bertolini, P.~Di Vecchia, M.~Frau, A.~Lerda and R.~Marotta,
``N = 2 gauge theories on systems of fractional D3/D7 branes,''
Nucl.\ Phys.\ B {\bf 621} (2002) 157
[arXiv:hep-th/0107057].

\bibitem{Bertolini:2001gq}
M.~Bertolini, P.~Di Vecchia and R.~Marotta,
``N = 2 four-dimensional gauge theories from fractional branes,''
arXiv:hep-th/0112195.

\bibitem{Gross:1998gk}
D.~J.~Gross and H.~Ooguri,
``Aspects of large N gauge theory dynamics as seen by string theory,''
Phys.\ Rev.\ D {\bf 58} (1998) 106002
[arXiv:hep-th/9805129].

\bibitem{Zarembo:1999bu}
K.~Zarembo,
``Wilson loop correlator in the AdS/CFT correspondence,''
Phys.\ Lett.\ B {\bf 459} (1999) 527
[arXiv:hep-th/9904149].

\bibitem{Olesen:2000ji}
P.~Olesen and K.~Zarembo,
``Phase transition in Wilson loop correlator from AdS/CFT correspondence,''
arXiv:hep-th/0009210.

\bibitem{Kim:2001td}
H.~Kim, D.~K.~Park, S.~Tamarian and H.~J.~Muller-Kirsten,
``Gross-Ooguri phase transition at zero and finite temperature: Two  circular Wilson loop case,''
JHEP {\bf 0103} (2001) 003
[arXiv:hep-th/0101235].

\bibitem{Zarembo:2001jp}
K.~Zarembo,
``String breaking from ladder diagrams in SYM theory,''
JHEP {\bf 0103}, 042 (2001)
[hep-th/0103058].

\bibitem{Wess}
J.Wess and J.~Bagger, {\it Supersymmetry and Supergravity} 
(Princeton University Press, 1982).

\end{thebibliography}
\end{document}